\documentclass[superscriptaddress,twocolumn,aps,prb,floatsfix]{revtex4}
\usepackage{graphicx}
\usepackage{amssymb,amsmath}
\usepackage{array}
\usepackage{dcolumn}

\begin{document}

 \title{The crucial importance of the $t_{2g}$--$e_g$ hybridization in 
 transition metal oxides}

\author{Sylvain Landron}
\author{Marie-Bernadette Lepetit}

\affiliation{CRISMAT, ENSICAEN-CNRS UMR6508, 6~bd. Mar\'echal Juin, 14050 Caen,
FRANCE}

\date{\today}
\begin{abstract}
We studied the influence of the trigonal distortion of the regular
octahedron along the (111) direction, found in the $\rm CoO_2$
layers. Under such a distortion the $t_{2g}$ orbitals split into one
$a_{1g}$ and two degenerated $e_g^\prime$ orbitals. We focused on the
relative order of these orbitals. Using quantum chemical calculations
of embedded clusters at different levels of theory, we analyzed the
influence of the different effects not taken into account in the
crystalline field theory~; that is metal-ligand hybridization, long-range
crystalline field, screening effects and orbital relaxation.  We found
that none of them are responsible for the relative order of the
$t_{2g}$ orbitals. In fact, the trigonal distortion allows a mixing of
the $t_{2g}$ and $e_g$ orbitals of the metallic atom. This 
hybridization is  at the origin of the
$a_{1g}$--$e_g^\prime$ relative order and of the incorrect 
prediction of the crystalline field theory.

\end{abstract}
\maketitle

\section{Introduction}
Since the discovery of super-conductivity in the hydrated $\rm
Na_{0.35} Co O_2-1.3H_2O$~\cite{supra} compound and of the very large
thermopower in the $\rm Na_{0.7\pm \delta} Co O_2$~\cite{TEP} members
of the same family, the interest of the community in systems built
from $\rm CoO_2$ layers has exploded.  The first step in the
understanding of the electronic properties of transition metal oxides,
such as the $\rm Co O_2$-based compounds, is the analysis of the
crystalline field splitting of the $d$ orbitals of the transition metal
atom. Indeed, depending on this splitting,  the
spin state of the atom, the nature of the Fermi level orbitals, and
thus the Fermi level properties will differ.

\begin{figure}[h]
\resizebox{!}{6cm}{\includegraphics{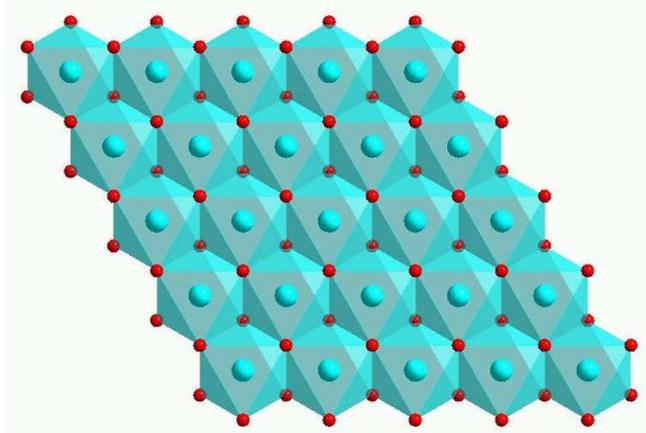}} 
\caption{Schematic representation of the $\rm CoO_2$ layers.}
\label{f:CoO2}
\end{figure}

The $\rm CoO_2$ layers are built from edge-sharing $\rm CoO_6$
octahedra (see figure~\ref{f:CoO2}).
In these layers, the first coordination shell of the metal atom
differs from the regular octahedron by a trigonal distortion along the
three-fold (111) axis (see figure~\ref{f:dist}).  In all known
materials (whether cobalt oxides or other metal oxides such as $\rm
LiVO_2$, $\rm NaTiO_2$, $\rm NaCrO_2$, etc\dots), this distortion is
in fact a compression.  The local
symmetry group of the metal atom is lowered from $O_h$ to
$D_{3d}$. The $T_{2g}$ irreducible representation of the $O_h$ group
is thus split into one $E_g$ and one $A_{1g}$ representations.  The
relative energies of the resulting $e_g^\prime$ and $a_{1g}$ orbitals
(see figure~\ref{f:dist}) has been a subject of controversy in the
recent literature, as far as the low spin $\rm Co^{4+}$ ion is
concerned. At this point let us point out the crucial importance of
the knowledge of this energetic order for the understanding of the low
energy properties of the $\rm CoO_2$ layers. Indeed, the possible
existence of an orbital order, as well as the minimal model pertinent
for the description of these systems depend on this order.

Authors such as Maekawa~\cite{Maekawa}, following the crystalline field
theory, support that the $a_{1g}$ orbital is of lower energy than the
two degenerated $e_g$ ones, leading to an orbital degeneracy for the
$\rm Co^{4+}$ ion. On the contrary, ab initio calculations, both using
periodic density functional methods~\cite{dft} and local quantum
chemical methods for strongly correlated systems~\cite{CoO2_1} yield
an $a_{1g}$ orbital of higher energy than the $e_g^\prime$ ones, and a non
degenerated Fermi level of the $\rm Co^{4+}$ ion.
\begin{figure}[h]
\resizebox{!}{6cm}{\includegraphics{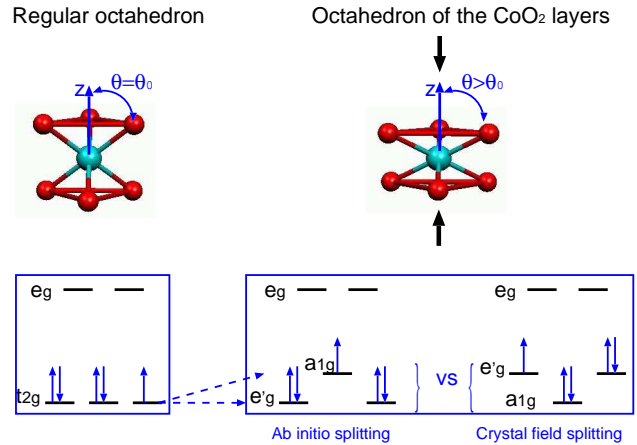}} 
\caption{Schematic representation of cobalt 3d splitting. $\theta$
represents the angle between the $\bf z$ axis ---~the 3-fold (111)
axis of the $\rm CoO_6$ octahedron~--- and the $\rm Co-O$
direction. $\theta_0=arccos{\left(\frac{1}{\sqrt{3}}\right)}\simeq
54.74^\circ$ is the $\theta$ angle for the regular octahedron.}
\label{f:dist}
\end{figure}
%
Angle Resolved Photoemission Spectroscopy (ARPES) experiments were
performed on several $\rm CoO_2$ compounds~\cite{arpes}.  This
technique probes the Fermi surface and clearly shows that the Fermi
surface of the $\rm CoO_2$ layers is issued from the $a_{1g}$
orbitals, and not at all from the $e_g^\prime$ orbitals (orbitals of
$E_g$ symmetry, issued from the former $t_{2g}$ orbitals), supporting
the ab-initio results.

In the present work, we will try to understand the reasons why the
crystalline field model is unable to find the good energetic order of
$t_{2g}$ orbitals in such trigonal distortions.  Several hypotheses
can be made to explain the orbital order~: the delocalization of the
metal $3d$ orbitals toward the ligands, the fact that the
electrostatic potential of the whole crystal differs from the one
assumed in the crystalline field model, the correlation effects within
the $3d$ shell, the screening effects, etc. All these hypotheses will
be specifically tested on the $\rm Co^{4+}$ ($3d^5$) ion that is subject in
this work to a more thorough study than other metal fillings. Nevertheless,
other metal fillings ($3d^1$ to $3d^3$, that can be found in vanadium,
titanium chromium, \dots oxides) will also be studied.  We will see
the crucial importance of the band filling on the $t_{2g}$ orbitals
order. In this work we will focus only on the $O_h$ to $D_{3d}$
trigonal distortion, subject of the controversy.

The next section will present the method used in this work, section
three and four will reports the calculations and analyze them, finally
the last section will be devoted to the conclusion.

\section{Computational method and details}
The energy of the atomic $3d$ orbitals is an essentially local value,
as supposed in the crystalline field model. However its analysis exhibits
some non local contributions. Indeed, orbitals energies can be seen as
resulting from the following terms: \begin{itemize}
\item the electrostatic potential due to the first coordination shell
  ---~in the present case, the six oxygen atoms of the octahedron,
  further referred as nearest neighbor oxygens (NNO)~---,
\item the electrostatic potential due to the rest of the crystal,
\item the kinetic energy that includes the hybridization of the metal
orbitals with nearest neighbor ligands, 
\item the Coulomb and exchange contributions within the $3d$ shell, 
\item the radial relaxation of the $3d$ orbitals,
\item and finally the virtual excitations from the other orbitals that
are responsible for the screening effects.
\end{itemize} 
All these contributions, excepts for the electrostatic potential due to
the rest of the crystal (nucleus attractions and 
Coulomb interactions), are essentially local contributions~\cite{revue}
and known to decrease very rapidly with the distance to the metal
atom. In fact, they are mostly restricted to the first coordination
shell of the cobalt.  On the contrary, the Madelung potential 
retains the resulting non local contributions from the nucleus
attraction and the Coulomb electron-electron repulsion. It is known to be
very slowly convergent with the distance. We thus made calculations at
different levels, including first all the above effects, and then
excluding them one at the time, in order to end up with the sole
effects included in the crystalline field model.

The calculations will thus be done on $\rm CoO_6$ or $\rm Co$
fragments. Different embedding and different levels of calculation
will be used. The $\rm Co-O$ distance will be fixed to the value of the
super-conducing compound, i.e. $R_{\rm Co-O} = 1.855$~\AA.  The angle
$\theta$ between the $\rm Co-O$ direction and the {\bf z} axis (see
figure~\ref{f:dist}) will be varied from 0 to $90^\circ$.

The calculations will be done at the Complete Active Space Self
Consistent Field~+~Difference Dedicated Configurations
Interaction~\cite{CASSCF,DDCI} (CASSCF+DDCI, see
subsection~\ref{ss:CAS+DDCI}) level for the most involved case, using
the core pseudopotential and basis set of Barandiaran {\em et
al.}~\cite{bases}. The fragment used will include all the first
coordination oxygens in addition to the cobalt atom. The embedding
will be designed so that to properly represent the full Madelung
potential of the super-conducting material, and the exclusion effects
of the rest of the crystal on the computed fragment electrons (see
reference~\cite{CoO2_1} for further details).
For the simplest case a minimal basis set derived from the preceeding
one will be used and only the cobalt atom will be included in the
computed fragment. The effect of the crystalline field will be described
by $-2$ point charges located at the positions of the first coordination
shell oxygens. The calculations will be done at the CASSCF level only.
Between these two extreme cases, several intermediate ones will be
considered, in order to check the previously enumerate points.

The electrostatic potential due to the cobalt first oxygen neighbors
(NNO), as well as the unscreened Coulomb and exchange contributions
within the $3d$ shell, are included in all calculations. The
electrostatic potential is treated either through the inclusion of the
NNO in the computed fragment or through $-2$ point charges. The
Coulomb and exchange contributions are treated through the CASSCF
calculation. The electrostatic contribution of the rest of the crystal
is included only in the most involved calculations, using an
appropriated embedding of point charges and Total Ions
pseudo-Potential~\cite{TIP}.  The hybridization of the metal $3d$
orbitals is treated by including explicitely the NNO in the considered
fragment ($\rm CoO_6$). The radial relaxation of the $3d$ orbitals is
treated when extended basis set are used. When a minimal basis set is
used, the radial part of the orbitals is frozen as in the high spin
state of the isolated $\rm Co^{4+}$ ion. Finally, the screening effects
are treated only when the calculation is performed at the CASSCF+DDCI
level.

\subsection{The CASSCF and DDCI methods}
\label{ss:CAS+DDCI}
Let us now described shortly the CASSCF and DDCI ab initio methods.
These methods are configurations interaction (CI) methods, that is
exact diagonalization methods within a selected set of Slater's
determinants. These methods were specifically designed to treat
strongly correlated systems, for which there is no qualitative
single-determinant description. The CASSCF method treats exactly all
correlation effects and exchange effects within a selected set of
orbitals (here the $3d$ shell of the cobalt atom). The DDCI method
treats in addition the excitations responsible for the screening
effects on the exchange, repulsion, hopping, etc.  integrals.  These
methods are based on the partitioning of the fragment orbitals into
three sets\begin{description}
\item[the occupied orbitals ] that are always doubly-occupied in all
determinants of the Complete Active Space or CAS (here the cobalt
inner electrons and the NNO ones),
\item[the active orbitals ] that can have all possible occupations and
spins in the CAS (here the cobalt $3d$ orbitals),
\item[the virtual orbitals ] that are always empty in the CAS. 
\end{description}
The CASCI method is the exact diagonalization within the above defined
Complete Active Space. The CASSCF method optimizes in addition the
fragment orbitals in order to minimize the CASCI wave function
energy. This is a mean-field method for the occupied orbitals but all
the correlation effects within the active orbitals are taken into
account. Finally the DDCI method uses a diagonalization space that
includes the CAS, all single- and double-excitations on all
determinants of the CAS, except the ones that excite to occupied
orbitals into two virtual orbitals. Indeed, such excitations can be
shown not to contribute ---~at the second order of perturbation~--- to the
energy differences between  states that differ essentially by their
CAS wave function. Therefore, they  have little importance for the present
work. The DDCI method thus accurately treats both the correlation
within the CAS and the screening effects.

Compared to the very popular density functional methods, the CAS+DDCI
method presents the advantage of treating exactly the correlation
effects within the $3d$ shell. This is an important point for strongly
correlated materials such as the present ones. Indeed, even if the DFT
methods should be exact provided the knowledge of the correct
exchange-correlation functional, the present functionals work very
well for weakly correlated systems, but encounter more difficulties
with strong correlation effects. For instance the LDA approximation
finds most of the sodium cobaltites compounds ferromagnetic~\cite{dft} in
contradiction with experimental results. LDA+U functionals try to
correct these problems by using an ad hoc on-site repulsion, U, within
the strongly correlated shells. This correction yields better results,
however it treats the effect of the repulsion within a mean field
approximation, still lacking a proper treatment of the strong
correlation. The drawbacks of the CAS+DDCI method compared to
the DFT methods are its cost in term of CPU time and necessity to work
on formally finite and relatively small systems. In the present case
however, this drawback appear to be an advantage since it decouples
the local quantities under consideration from the dispersion problem.

\section{Results and Analysis}

\begin{figure}[h]
\resizebox{4.5cm}{!}{\includegraphics{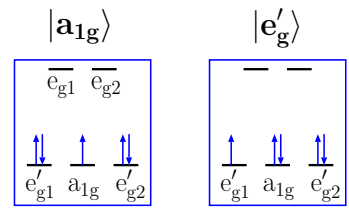}} 
\caption{Schematic representation of the $\rm Co^{4+}$ states of
interest. Let us point out that ${\rm \bf |e_g^\prime\rangle}$ is
doubly-degenerated, the hole being located either on the
$e_{g1}^\prime$ or on the $e_{g2}^\prime $ orbitals.}
\label{f:orb_eff}
\end{figure}

Let us first attract the attention of the reader on what is supposed
to be the energy difference between the $e_g^\prime$ and $a_{1g}$
orbitals of the $\rm Co^{4+}$ ion in an effective model.  In fact, the
pertinent parameters for an effective model should be such that one
can reproduce by their means the exact energies or, in the present
case, the ab-initio calculation of the different $\rm Co^{4+}$ atomic
states. It results, that within a Hubbard type model, the pertinent
effective orbital energies should obey the following set of equations 
\begin{eqnarray*}
E\left({\rm \bf |a_{1g}\rangle}\right) &=& 4 \varepsilon(e_g^\prime) + 
\varepsilon(a_{1g}) + 2U + 8U^\prime - 4 J_H \\
E\left({\rm \bf |e_{g}^\prime\rangle }\right) &=&  3 \varepsilon(e_g^\prime) + 
2 \varepsilon(a_{1g}) + 2U + 8U^\prime - 4 J_H \\
\Delta E &=&E\left({\rm \bf |e_g^\prime\rangle} \right) -  E\left({\rm \bf |a_{1g}\rangle}\right)\\
&=& 
\varepsilon(a_{1g}) - \varepsilon(e_g^\prime) 
\end{eqnarray*}
where the schematic picture of the ${\rm \bf |e_{g}^\prime\rangle} $ and
${\rm \bf |a_{1g}\rangle}$ states is given in figure~\ref{f:orb_eff},
$\varepsilon(e_g^\prime)$ and $\varepsilon(a_{1g})$ are the effective
orbital energies of the $e_g^\prime$ and $a_{1g}$ atomic orbitals,
$U$ is the effective electron-electron repulsion of two electrons in
the same cobalt $3d$ orbital, $U^\prime$ the effective repulsion of
two electrons in different cobalt $3d$ orbitals and $J_H$ the atomic
Hund's exchange effective integrals within the cobalt $3d$ shell.

\subsection{The reference calculation}
The reference calculation includes all effects detailed in the
preceding section. For the super-conducting compound the effective
$t_{2g}$ splitting was reported in reference~\cite{CoO2_1} to be 
$$ \Delta E =  \varepsilon(a_{1g}) - \varepsilon(e_g^\prime) = 315~\rm meV $$
 This point corresponds to $\theta \simeq 61.5^\circ$ (that is a value
of $\theta$ larger than the one of the regular octahedron
$\theta_0\simeq 54.74^\circ$) where the crystalline field theory predicts
a reverse order between the $t_{2g}$ orbitals.

\subsection{Screening effects}
The effect of the screening on the $t_{2g}$ orbital splitting can be
evaluated by doing a simple CASCI calculation using the same fragment,
embedding, basis set and orbitals as the preceding calculation. 
Without the screening effects, one finds a $t_{2g}$ splitting of
$$ \Delta E = \varepsilon(a_{1g}) - \varepsilon(e_g^\prime) = 428~\rm meV $$ 
Obviously the screening effects cannot be taken as responsible
for the qualitative energetic order between the $a_{1g}$ and
$e_g^\prime$ orbitals. 

\subsection{Cobalt $3d$~--~oxygen hybridization}
The effect of the hybridization of the cobalt $3d$ orbitals with the
neighboring oxygen ligands can be evaluated by taking out the oxygen
atoms from the quantum cluster, and treating them as simple $-2$ point
charges at the atomic locations. The other parameters of the
calculation are kept as in the preceding case. The new orbitals are
optimized at the average-CASSCF level between the two
${\rm \bf |e_g^\prime\rangle}$ and the ${\rm \bf |a_{1g}\rangle}$ states. 
It results in a $t_{2g}$ splitting of  
$$ \Delta E = \varepsilon(a_{1g}) - \varepsilon(e_g^\prime) = 40~\rm meV $$ 
for the super-conducting compound.  Again the hybridization of the
cobalt $3d$ orbitals with the neighboring oxygens cannot be taken as
responsible for the inversion of the splitting between the $a_{1g}$
and $e_g^\prime$ orbitals.

\subsection{Long-range electrostatic potential}
The effect of the long-range electrostatic potential can be evaluated by
restricting the embedding to the NNO point charges only, that is to the
electrostatic potential considered in the crystalline field method.
One finds a  $t_{2g}$ splitting of  
$$ \Delta E = \varepsilon(a_{1g}) - \varepsilon(e_g^\prime) = 124~\rm meV $$ 
Once again  the results is positive and thus the long-range
electrostatic potential is not the cause of the crystalline field
inversion of the $t_{2g}$ splitting.

\subsection{Orbital radial relaxation}
\label{ss:e}
At this point only few effects on top of the crystalline field theory are
still treated in the calculation. One of them is the radial
polarization effect of the $3d$ orbitals, that allows their adaptation
to the different occupations in the specific ${\rm \bf
|a_{1g}\rangle}$ and ${\rm \bf |e_g^\prime \rangle}$ states.  This
polarization is due to the use of an extended basis set. We thus
reduce the basis set to a minimal basis set (only one orbital degree
of freedom per $(n,l)$ occupied or partially occupied atomic shell).
The minimal basis set was obtained by the contraction of the extended
one~; the radial part of the orbitals being frozen as the one of the
the isolated $\rm Co^{4+}$ high spin state. This choice was done in
order to keep a basis set as close as possible to the extended one, and
because only for the isolated atom all $3d$ orbitals are equivalent,
and thus have the same radial part.  One obtains in this minimal basis
set a $t_{2g}$ splitting of
$$ \Delta E = \varepsilon(a_{1g}) - \varepsilon(e_g^\prime) = 41~\rm meV $$ 
At this point we computed the effective orbital energies in the sole
crystalline field conditions, however the result is still reverse than
what is usually admitted within this approximation. Indeed, the $\rm
Co^{4+}$ ion was computed in the sole electrostatic field of the NNO, 
treated as $-2$ point charges, the calculation is done within a
minimal basis set, and at the average-CASSCF level.

\subsection{Further analysis}
\label{ss:f}
\begin{figure}[h]
\resizebox{!}{6cm}{\includegraphics{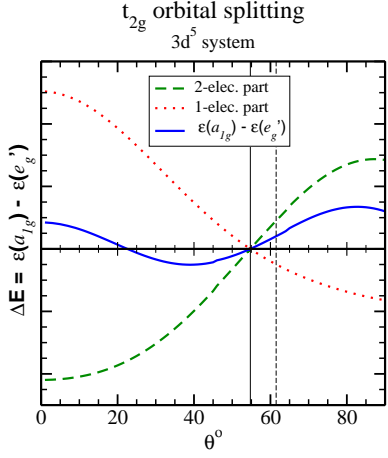}} 
\caption{Orbital splitting between the $a_{1g}$ and $e_g^\prime$
orbitals when only the nearest neighbor ligands electrostatic field is
included. The dotted red curve corresponds to the single-electron part
of the orbital energy difference~: $\Delta E_1$, that is the kinetic
energy (equation~(\ref{eq:T})), the electron-charge interaction
(equation~(\ref{eq:N})) and the interaction with the core electrons
(equation~(\ref{eq:occ})) . The dashed green curve corresponds to the
two-electron part of the orbital energy difference~: $\Delta E_2$,
that is the repulsion and exchange terms within the $3d$ shell
(equation~(\ref{eq:3d})). The solid vertical line points out the regular
octahedron $\theta$ value and the dashed vertical line the $\theta$
value for the super-conducting compound.}
\label{f:deltaE}
\end{figure}

In order to understand this puzzling result, we plotted the whole
curve $\Delta E (\theta)$ (see figure~\ref{f:deltaE}) at this level of
calculation and analyzed separately all energetic terms involved in
this effective orbital energy difference.

One sees on figure~\ref{f:deltaE} that the $\Delta E(\theta)$ curve is
not  monotonic, as expected from the crystalline field theory. Indeed,
while for $\theta=0$ the relative order between the $a_{1g}$ and
$e_g^\prime$ orbitals is in agreement with the crystalline field
predictions, for $\theta=90^\circ$ the order is reversed. One should
also notice that, in addition to the $\theta_0$ value of the regular
octahedron, there is another value of $\theta$ for which the three
$t_{2g}$ orbitals are degenerated. In the physically realistic region
of the trigonal distortion (around the regular octahedron $\theta_0$
value) the relative order between the $a_{1g}$ and $e_g^\prime$
orbitals is reversed compared to the crystalline field predictions.

Let us now decompose $\Delta E (\theta)$ into 
\begin{itemize}
\item its two-electron part within the $3d$ shell ---~$\Delta
E_2(\theta)$~---
\item and the rest referred as $3d$ single-electron part ---~$\Delta
E_1(\theta)$. $\Delta E_1$ includes the kinetic energy, the
electron-nucleus and electron-charge interaction, and the interaction
of the $3d$ electrons with the inner shells electrons.
\end{itemize}
One thus has
\begin{eqnarray*} 
\Delta E &=& \Delta E_1 + \Delta E_2 \\
&=& \varepsilon(a_{1g}) - \varepsilon(e_{g1}^\prime) 
  = \varepsilon(a_{1g}) - \varepsilon(e_{g2}^\prime) 
\end{eqnarray*}
with {\small
\begin{eqnarray} \hspace*{-2.0cm} 
\Delta E_1 &=& \quad
  \left\langle a_{1g}\left| -\frac{\nabla^2}{2} \right| a_{1g} \right\rangle 
\; - \;
  \left\langle e_g^\prime\left| -\frac{\nabla^2}{2} \right| e_g^\prime \right\rangle 
\label{eq:T} \\&&
+  \left\langle  a_{1g}  \left|\sum_{N} \frac{-Z_N}{R_N} \right|a_{1g}\right\rangle 
-  \left\langle  e_g^\prime  \left|\sum_{N} \frac{-Z_N}{R_N}\right|e_g^\prime \right\rangle  \label{eq:N} \\&&
+ \sum_{\chi~:~occ}
   2 \left\langle  a_{1g}\,\chi \left|\frac{1}{r_{12}}\right|a_{1g}\,\chi \right\rangle  
   - \left\langle a_{1g}\,\chi \left|\frac{1}{r_{12}}\right|\chi\,a_{1g}\right\rangle \nonumber \\&&
-  \sum_{\chi~:~occ} 
   2 \left\langle e_g^\prime\,\chi \left|\frac{1}{r_{12}}\right|e_g^\prime\,\chi \right\rangle  
   - \left\langle e_g^\prime\,\chi \left|\frac{1}{r_{12}}\right|\chi\,e_g^\prime \right\rangle \label{eq:occ}
\end{eqnarray} }
and {\small
\begin{eqnarray} \hspace*{-2.5cm} 
\Delta E_2 &=& \quad 
 \left\langle  a_{1g}\,a_{1g}\left|\frac{1}{r_{12}}\right| a_{1g}\,a_{1g} \right\rangle 
-\left\langle e_g^\prime \,e_g^\prime\left|\frac{1}{r_{12}}\right| e_g^\prime\,e_g^\prime \right\rangle  
\nonumber \\&&
+ 2\left\langle a_{1g}\,e_g^\prime \left|\frac{1}{r_{12}}\right|a_{1g}\, e_g^\prime \right\rangle
  -\left\langle a_{1g}\, e_g^\prime \left|\frac{1}{r_{12}}\right|e_g^\prime\, a_{1g}  \right\rangle  
\label{eq:3d}  \\&&
-  2\left\langle e_{g1}^\prime\, e_{g2}^\prime \left|\frac{1}{r_{12}}\right|e_{g1}^\prime\, e_{g2}^\prime\right\rangle
  +\left\langle e_{g1}^\prime\, e_{g2}^\prime \left|\frac{1}{r_{12}}\right|e_{g2}^\prime\, e_{g1}^\prime \right\rangle 
  \nonumber 
\end{eqnarray} }
where the equations are given in atomic units.  $Z_N$ refers to the
nucleus charge of the cobalt atom and the $-2$ point charges located
at the NNO positions. $R_N$ is the associated electron-charge
distance. The sum on $\chi$ runs over all the orbitals of the cobalt
inner-shells. 

Let us now examine the dependence on $\theta$ of each of the terms of
$\Delta E_1$ and $\Delta E_2$.
\begin{description}
  \item[Kinetic energy~:] the radial part of each of the $3d$ orbitals
  being identical due the the minimal basis set restriction, the
  kinetic part is identical for all $3d$ orbitals and thus its
  contribution to $\Delta E_1$ (terms labeled~\ref{eq:T} of $\Delta
  E_1$) vanishes.

  \item[Nuclear interaction~:] obviously this contribution to $\Delta
  E_1$ (terms labeled~\ref{eq:N} of $\Delta E_1$) strongly depends on
  $\theta$ through the position of the $-2$ charges.

  \item[Interaction with the inner-shells electrons~:] this term
  (terms labeled~\ref{eq:occ} of $\Delta E_1$) depends only on the
  shape of the $t_{2g}$ and inner-shells orbitals. However, the
  minimal basis set does not leave any degree of freedom for the
  relaxation of the inner-shells orbital whose shapes are thus
  independent of $\theta$. Similarly, the $3d$ radial part of the $3d$
  orbitals is totally frozen.

  \item[$\bf \Delta E_2$~:] finally, the dependence of $\Delta E_2$ can
  only go through the shape of the $a_{1g}$ and $e_g^\prime$ orbitals
  whose radial part is totally frozen due to the use of a minimal
  basis set.
\end{description}
If one accepts that the $a_{1g}$ and $e_g^\prime$ orbitals are issued
from the $t_{2g}$ orbitals of the regular octahedron, their angular form is
totally given by the symmetry (see eq.~\ref{eq:eg},~\ref{eq:t2g})
and both $\Delta E_2$ and the third contribution of $\Delta E_1$
should be independent of $\theta$.
\begin{equation}
  e_g  \left\{ 
  \begin{array}{ccl}
    e^{\circ}_{g1} & = & \frac{1}{\sqrt{3}}d_{xy}     
         +\frac{\sqrt{2}}{\sqrt{3}}d_{xz} \\[+1.5ex]
    e^{\circ}_{g2} & = & \frac{1}{\sqrt{3}}d_{x^2-y^2}
         +\frac{\sqrt{2}}{\sqrt{3}}d_{yz}
  \end{array} \right.  \label{eq:eg}
\end{equation}
\begin{equation}
  t_{2g} \left\{ 
  \begin{array}{ccl}
    a^{\circ}_{1g} & = & d_{z^2} \\[+1.5ex]
    e^{\circ \prime}_{g1} & = & \frac{\sqrt{2}}{\sqrt{3}}d_{xy}
        -\frac{1}{\sqrt{3}}d_{xz} \\[+1.5ex]
    e^{\circ \prime}_{g2} & = & \frac{\sqrt{2}}{\sqrt{3}}d_{x^2-y^2}
        -\frac{1}{\sqrt{3}}d_{yz}
  \end{array} \right.  \label{eq:t2g}
\end{equation}
where the $x$, $y$ and $z$ coordinates are respectively associated with
the $\bf a$, $\bf b$ and $\bf c$ crystallographic axes.

\begin{figure}[h]
\resizebox{!}{6cm}{\includegraphics{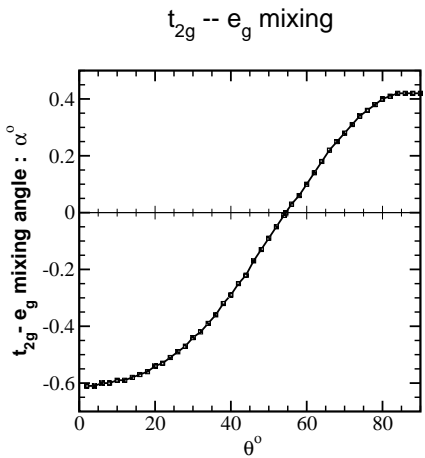}} 
\caption{$t_{2g}$--$e_g$ hybridization angle under the trigonal distortion.}
\label{f:hyb}
\end{figure}

Figure~\ref{f:deltaE} displays both $\Delta E_1$ (dotted red curve)
and $\Delta E_2$ (dashed green curve) contributions to $\Delta E$. One
sees immediately that $\Delta E_2$ is not at all independent of
$\theta$ but rather monotonically increasing with $\theta$. It results
that the above hypotheses of the $t_{2g}$ exclusive origin for the
$e_g^\prime$ orbitals is not valid. Indeed, out of the
$\theta=\theta_0$ point, the only orbital perfectly defined by the
symmetry is the $a_{1g}$ orbital. The $e_g^\prime$ and $e_g$ orbitals
belong to the same irreducible representation ($E_g$) and can thus mix
despite the large $t_{2g}$--$e_g$ energy difference. If we name this
mixing angle $\alpha$, it comes
$$
\begin{array}{ccccl}
  e_{gi} & = &  \quad e_{gi}^{\circ \prime} \cos{\alpha}&+&
                e_{gi}^{\circ} \sin{\alpha} \\[+1.5ex] 
   e_{gi}^\prime & =& - e_{gi}^{\circ \prime} \sin{\alpha} &+& 
                e_{gi}^{\circ} \cos{\alpha}
\end{array} 
$$
Figure~\ref{f:hyb} displays $\alpha$ as a function of $\theta$.
One sees that the $t_{2g}$--$e_g$ hybridization angle $\alpha$ is non
null ---~except for the regular octahedron~--- and a monotonic,
increasing function of $\theta$. Even if very small ($\pm 0.6^\circ$),
this $t_{2g}$--$e_g$ hybridization has an important energetic effect,
since it lowers the the $e_g^\prime$ orbital energy while increasing
the $e_g$ one. $\alpha$ is very small but it modulates large energetic
factors in $\Delta E_2$~: on-site Coulomb repulsions of two electrons
in the $3d$ orbitals. The result is a monotonic increasing variation
of $\Delta E_2$ as a function of $\theta$. The variation of the
$\Delta E_1$ term is dominated by its nuclear interaction part and
exhibits a monotonic decreasing variation as a function of $\theta$,
as expected from the crystalline field theory. The nuclear interaction and
$t_{2g}$--$e_g$ hybridization have thus opposite effects on the
$a_{1g}$--$e_g^\prime$ splitting. {\bf The failure of the crystalline field
theory thus comes from not considering the $t_{2g}$--$e_g$
hybridization.}

In the calculations presented in figures~\ref{f:deltaE} and
\ref{f:hyb}, the screening effects on the on-site Coulomb repulsions
and exchange integrals were not taken into account. Thus, the absolute
value of $\Delta E_2$ as a function of the hybridization $\alpha$, is
very large and $\alpha$ is very small. When the screening effects are
properly taken into account, the absolute value of $\Delta E_2$ as a
function of $\alpha$ is reduced by a factor about 6, and the
$t_{2g}$--$e_g$ hybridization is much larger than the values presented
in figure~\ref{f:hyb}. Indeed, in the superconducting compound, for a
realistic calculation including all effects, one finds $\alpha\simeq
13^\circ$ ($\theta=61.5^\circ$).

At this point we would like to compare the $a_{1g}$--$e_g^\prime$
splitting found in the present calculations and the one found using
DFT methods. Indeed, our splitting (315~meV for the superconducting
compound) is larger than the DFT evaluations (always smaller $<
150~\rm meV$). This point can be easily understood using the
single-electron and two-electron part analysis presented
above. Indeed, while the single-electron part is perfectly treated in
DFT calculations, the two-electron part is treated within the
exchange-correlation kernel. However these kernels are well known to
fail to properly reproduce the strong correlation effects present in
the transition metal opened $3d$ shells. One thus expect that while the
single-electron part of the atomic orbital energies is well treated,
the two-electron part is underestimated, resulting in an
under-evaluation of the $a_{1g}$--$e_g^\prime$ splitting, as can be
clearly seen from figure~\ref{f:deltaE}.

\section{Other cases}

We considered up to now a $\rm Co^{4+}$ ion, that is five electrons in
the $3d$ shell, and a fixed metal--ligand distance, $R_{\rm M-O}$.
Let us now examine the effect of the distance $R{\rm M-O}$ and the
band filling on the $a_{1g}$--$e_g^\prime$ splitting.  The
calculations presented in this section follow the same procedure as in
sections~\ref{ss:e},~\ref{ss:f}. For different fillings a typical
example in the transition metal oxides family was used to define the
type of metallic atom and metal oxygen distances. Minimal basis set
issued from full contraction of the basis set given in
reference~\cite{bases} will be used.

\subsection{The effect of the Co--O distance}
Figure~\ref{f:dist} displays the $a_{1g}$--$e_g^\prime$ energy
splitting as a function of the distortion angle $\theta$ and for
different distances. The range of variation~: from 1.8\AA\ to 1.95\AA,
includes all physically observed distances in $\rm CoO_2$ layers. 
%
\begin{figure}[h]
\resizebox{!}{6cm}{\includegraphics{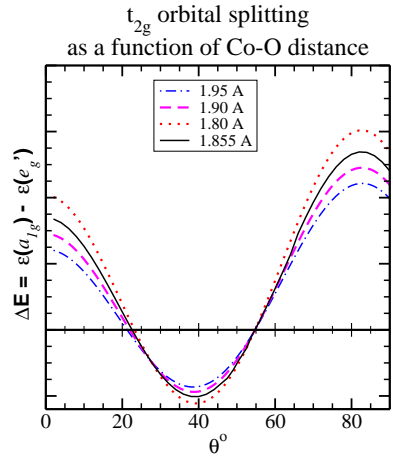}} 
\caption{Orbital splitting between the $a_{1g}$ and $e_g^\prime$
orbitals for a $3d^5$ transition metal and for different metal--ligand
distances.  Only the nearest neighbor ligands electrostatic field is
included in the calculation.
The dotted red curve corresponds to $R_{\rm Co-O}=1.8~\rm \AA$, the
solid black curve corresponds to the superconducting compound ($R_{\rm
Co-O}=1.855~\rm \AA$), the magenta dashed curve corresponds to
$R_{\rm Co-O}=1.9~\rm \AA$, and finally the dot-dashed blue curve
corresponds to $R_{\rm Co-O}=1.95~\rm \AA$.}
\label{f:dist}
\end{figure}
%
One sees immediately that despite the large variation of the
metal--ligand distance, the relative order of the $a_{1g}$ and
$e_g^\prime$ orbitals remains identical. The main effect of $R{\rm
M-O}$ is thus to renormalize the amplitude of the splitting, lowering
the splitting for larger distances and increasing it for smaller ones.

\subsection{$3d^1$}
The simplest filling case corresponds to only one electron in the $3d$
shell. This is, for instance, the case of the $\rm NaTiO_2$ compound.
The calculations were done using the average Ti--O distance found in
$\rm NaTiO_2$~\cite{85657}~: $R_{\rm Ti-O}=2.0749 \rm \AA$.

In this case, $\Delta E_2=0$ and $\Delta E(\theta)=\Delta E_1(\theta)$
behaves as pictured in figure~\ref{f:deltaE}. The $a_{1g}$ orbital is
of lower energy than the $e_g ^\prime$ for $\theta > \theta_0$ and of
higher energy for $\theta < \theta_0$. This result is in perfect
agreement with the crystalline field theory.

\subsection{$3d^2$}
\begin{figure}[h]
\resizebox{!}{6cm}{\includegraphics{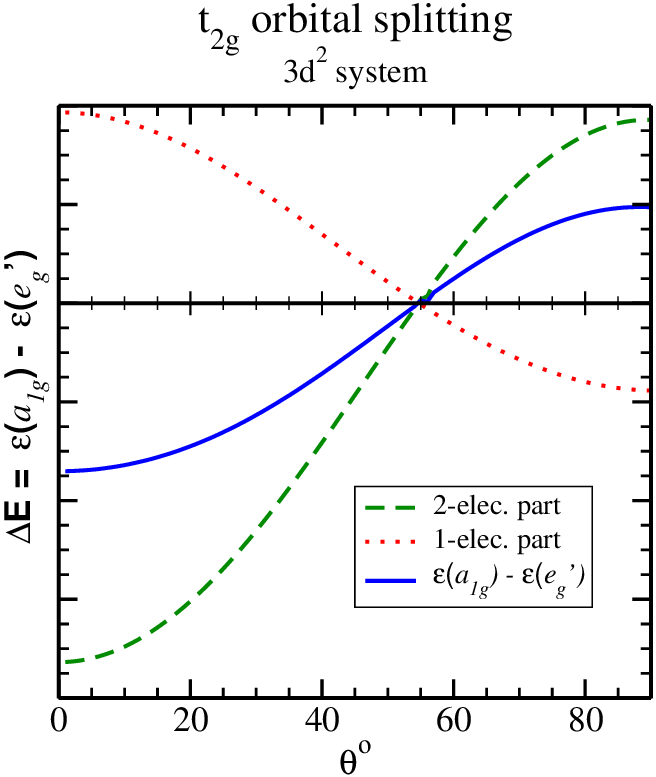}} 
\caption{Orbital splitting between the $a_{1g}$ and $e_g^\prime$
orbitals for a $3d^2$ transition metal. Only the nearest neighbor
ligands electrostatic field is included in the calculation. 
The dotted red curve corresponds to the single-electron part
of the orbital energy difference~: $\Delta E_1$, that is the kinetic
energy (equation~(\ref{eq:T})), the electron-charge interaction
(equation~(\ref{eq:N})) and the interaction with the core electrons
(equation~(\ref{eq:occ})) . The dashed green curve corresponds to the
two-electron part of the orbital energy difference~: $\Delta E_2$,
that is the repulsion and exchange terms within the $3d$ shell
(equation~(\ref{eq:3d})).}
\label{f:LiVO2}
\end{figure}

A simple example of the $3d^2$ filling in transition metal oxides is
the $\rm LiVO_2$ compound. Indeed, the vanadium atom is in the $\rm V^{3+}$
ionization state. We thus used a metal oxygen distance of
$R_{\rm V-O}=1.9787 \rm \AA$~\cite{202540}. 
 Figure~\ref{f:LiVO2} displays the $a_{1g}$--$e_g
^\prime$ splitting as well as its decomposition into the single-electron
and two-electron parts.
As in the $3d^5$ case (figure~\ref{f:deltaE}), the single-electron and
two-electron parts behave in a monotonic way as a function of $\theta$,
and in an opposite manner. In the present case, however, the
two-electron part always dominates over the one-electron part and the
$a_{1g}$--$e_g^\prime$ orbital splitting is always reversed compared
to the crystalline field predictions. As for the $3d^5$ system, there is a
slight $e_g^\prime$--$e_g$ hybridization that is responsible for the
$t_{2g}$ orbitals order.

\subsection{$3d^3$}
\begin{figure}[h]
\resizebox{!}{6cm}{\includegraphics{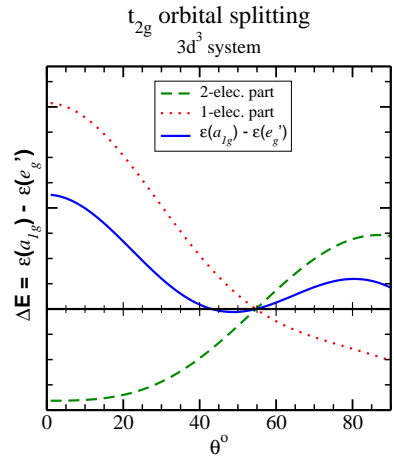}} 
\caption{Orbital splitting between the $a_{1g}$ and $e_g^\prime$
orbitals for a $3d^3$ transition metal. Only the nearest neighbor
ligands electrostatic field is included in the calculation. 
The dotted red curve corresponds to the single-electron part
of the orbital energy difference~: $\Delta E_1$, that is the kinetic
energy (equation~(\ref{eq:T})), the electron-charge interaction
(equation~(\ref{eq:N})) and the interaction with the core electrons
(equation~(\ref{eq:occ})) . The dashed green curve corresponds to the
two-electron part of the orbital energy difference~: $\Delta E_2$,
that is the repulsion and exchange terms within the $3d$ shell
(equation~(\ref{eq:3d})).}
\label{f:NaCrO2}
\end{figure}

Examples of $3d^3$ transition metal oxides are found easily in the
chromium compounds. Let us take for instance the $\rm NaCrO_2$
system~\cite{24595}.  The metal oxygen distance is thus~: $R_{ \rm Cr-O}
\simeq 1.901 \rm \AA$.  Figure~\ref{f:NaCrO2} displays the
$a_{1g}$--$e_g^\prime$ orbital splitting as well as its decomposition
into single- and two-electron parts.
As usual the single-electron part and the two-electron part are
monotonic as a function of $\theta$ but with slopes of opposite
signs. This case is quite similar to the $3d^5$ case since none of the
single- and two- electron parts dominates the $t_{2g}$ orbital
splitting over the whole range. Indeed, for small values of $\theta$,
the crystalline field effect dominates and the $a_{1g}$ orbital is above
the $e_g^\prime$ ones while, for large values of $\theta$, the
two-electron part dominates and the $a_{1g}$ orbital is again above
the $e_g^\prime$ ones. In a small intermediate region the order is
reversed.  In the realistic range of $\theta$ ($\theta \simeq
\theta_0$) there is a strong competition between the two effects
(quasi-degeneracy of the $a_{1g}$ and $e_g^\prime$ orbitals) and no
simple theoretical prediction can be made. The crystalline field theory is
not predictive but the present calculations cannot be considered as
predictive either, since all the neglected effects may reverse the
$a_{1g}$--$e_g^\prime$ order.

\section{Discussion and conclusion}

In the present work we studied the validity of the crystalline field
theory under the application of a trigonal distortion on the regular
octahedron. Under such a distortion, the $T_{2g}$
irreducible representation (irrep) of the $O_h$ group spits into
$A_{1g}$ and $E_g$ irreps ($T_{2g} \longrightarrow A_{1g} \oplus
E_g$), while the $e_g$ irrep remains untouched ($E_g \longrightarrow
E_g$). The hybridization between the $t_{2g}$ and $e_g$ orbitals thus become
symmetry allowed, even if hindered by energetic factors. This
hybridization is not taken into account in the crystalline field theory.
It is however of crucial importance for the relative order between the
former $t_{2g}$ orbitals and the reason of the failure of the crystalline
field theory to be predictive. Indeed, due to the $t_{2g}$--$e_g$
orbitals hybridization, the two-electron part of the $e_g^\prime$
orbital energy becomes dependant of the amplitude of the 
distortion and of opposite effect to the single-electron part. The
relative order of the $t_{2g}$ orbitals thus depends on the competition
between these two effects and as a consequence of the band filling.

In this work we studied the $O_h$ to $D_{3d}$ distortion, however
one can expect similar effects to take place for other distortions of
the regular octahedron. 
The condition for these effects to take place is that  
the $T_{2g}$ irreducible
representation splits into a one-dimensional irrep ($A$) and the same
two-dimensional irrep ($E$) as the one the $e_g$ orbitals are
transformed to 
\begin{eqnarray*}
T_{2g} &\longrightarrow & A \oplus E\\
E_g &\longrightarrow& E  
\end{eqnarray*} \hfill \\
Indeed, under such a distortion, $t_{2g}$--$e_g$ hybridization
phenomena are allowed. The distortion should thus transform $O_h$ into
sub-groups that keep the $C_3$ (111) symmetry axis~: $C_3$, $C_{3v}$,
$D_3$, $S_6$ and $D_{3d}$. Examples of such deformations are the
elongation of the metal--ligand distance of one of the sets of three
symmetry related ligands, or the rotation of such a set three ligands
around the (111) symmetry axis. For instance, one will expect that
$t_{2g}$--$e_g$ hybridization will also take place in trigonal
prismatic coordination.

However, in real systems like the sodium cobaltites, these distortion
do not usually appear alone but rather coupled. For instance, in the
squeezing of the metal layer between the two oxygen layers observed as
a function of the sodium content in $\rm Na_xCoO_2$, the Co--O bond
length and the three-fold trigonal distortion are coupled. Since this
composed distortion belongs to the above-cited class, the
$t_{2g}$--$e_g$ hybridization will take place and the relative orbital
order between the $a_{1g}$ and $e_g^\prime$ orbitals will be
qualitatively the same as in figure~\ref{f:deltaE}. The bond length
modification at equal distortion angle, $\theta$, will only change the
quantitative value of the orbital splitting, but not its sign. A bond
elongation reduces the splitting  a bond compression increases
it. One can thus expect in sodium cobaltites that the
$a_{1g}$--$e_g^\prime$ orbital energy splitting will decrease with
increasing sodium content. The reader should however have in mind that
the effects of this splitting reduction will remain relatively small
compared to the band width as clearly seen in
reference~\cite{picket}. In fact, one can expect that a large  effect
will be the modification of the band dispersion due not only to the
bond length modification, but also to the $t_{2g}$--$e_g$ hybridization.

\acknowledgments  The authors thank Jean-Pierre Doumerc and Michel
Pouchard for helpful discussions and Daniel Maynau for providing us
with the CASDI suite of programs. These calculations where done using
the CNRS IDRIS computational facilities under project n$^\circ$1842.


\end{document}